\documentclass[prc,aps,superscriptaddress,showpacs,nofootinbib,twocolumn]{revtex4}
\usepackage{amssymb}
\usepackage[dvips]{graphicx}
\usepackage[english]{babel}
\usepackage{indentfirst}
\usepackage{amsxtra}
\usepackage{amsmath}
\usepackage{supertabular}
\usepackage{multirow}
\usepackage[mathcal]{eucal}
\usepackage[usenames]{color}
\usepackage{ulem}

\begin{document}
\title {\bf Neutron $2p$ and $1f$ spin--orbit splittings in $^{40}$Ca, $^{36}$S, and $^{34}$Si $N=20$ isotones: tensor--induced and pure spin--orbit effects}

\author{M. Grasso}
\affiliation{Institut de Physique Nucl\'eaire, IN2P3-CNRS, Universit\'e Paris-Sud, 
F-91406 Orsay Cedex, France}

\author{M. Anguiano}
\affiliation{Departamento de Fisica At\'omica, Molecular y Nuclear, Universidad de Granada, E-18071 Granada, Spain}

\begin{abstract} 
Neutron $2p$ and $1f$ spin--orbit splittings 
 were recently measured in the isotones $^{37}$S and $^{35}$Si by $(d,p)$ transfer reactions. 
Values were reported by using the major fragments of the states. 
An important reduction of the $p$ splitting was observed, from 
 $^{37}$S to $^{35}$Si, associated to a strong modification of the spin--orbit potential in the central region of the nucleus $^{35}$Si.  

We analyze  $2p$ and $1f$ neutron spin--orbit splittings 
in the $N=20$ isotones $^{40}$Ca, $^{36}$S, and $^{34}$Si. 
We employ several Skyrme and Gogny interactions, to reliably  isolate pure spin--orbit and tensor--induced contributions, within the mean--field approximation.  We use interactions (i) without the tensor force; (ii) with the tensor force and with tensor parameters adjusted on top of existing 
parametrizations; (iii) with the tensor force and with tensor and spin--orbit parameters adjusted simultaneously on top of existing parametrizations. 
 
We predict in cases (ii) and (iii) a non negligible reduction of both $p$ and $f$ splittings, associated to neutron--proton tensor effects, from $^{40}$Ca to $^{36}$S. The two splittings are further decreased for the three types of interactions, going from $^{36}$S to $^{34}$Si. This reduction is produced by the   
spin--orbit force and is not affected by tensor--induced contributions. For both reductions, from $^{40}$Ca to $^{36}$S and from 
 $^{36}$S to $^{34}$Si,  
 we predict in all cases that the modification is more pronounced for $p$ than for  $f$ splittings.
The measurement of the centroids for neutron $2p$ and $1f$ states in the nuclei $^{36}$S and $^{34}$Si 
 would be interesting to validate this prediction experimentally. 
We show the importance of using interactions of type (iii), because they provide 
 $p$ and $f$ splittings in the nucleus $^{40}$Ca which are in agreement with the corresponding experimental values. 
 
\end{abstract} 

\vskip 0.5cm \pacs {21.60.Jz,21.10.Pc,21.30.Fe} \maketitle 
%

\section{Introduction}
We recently proposed parametrizations for the tensor mean--field force, introduced on top of existing Skyrme and Gogny interactions \cite{angu2012,gra2013}. 

In Ref. \cite{angu2012}, we investigated in particular the Gogny case. A few previous studies concerning the 
tensor force in the Gogny interaction existed in the literature \cite{otsu,co}. In both studies \cite{otsu,co}, the authors introduced a finite--range tensor--isospin term. The 
set proposed in Ref. \cite{otsu} was employed in Ref. \cite{moreno} to analyze neutron--proton tensor effects on some single--particle spectra. 
In Ref. \cite{angu2012}, we discussed the need of introducing not only a tensor--isospin term, but also a pure tensor term. This allowed us to adjust independently  
the neutron--proton and the like--particle tensor contributions in the Gogny case, as was already done with the Skyrme force.  

Since tensor and spin--orbit effects are strictly connected in modifying single--particle splittings along isotopic and isotonic chains, the importance of refitting simultaneously also the spin--orbit parameter was discussed in Ref. \cite{gra2013}, for both Skyrme and Gogny cases. A three--step procedure \cite{zale1,zale2} was employed by fitting the neutron $1f$ spin--orbit splitting in the nuclei $^{40}$Ca, $^{48}$Ca, and $^{56}$Ni. The first nucleus, which is spin--saturated, was used to adjust the spin--orbit parameter; 
the nucleus $^{48}$Ca, which is spin--saturated in protons, was used to adjust the like--particle tensor contribution;  
finally, the nucleus $^{56}$Ni, which is unsaturated, was used to adjust the neutron--proton contribution. This three--step procedure was useful for 
identifying reasonable signs and regions of values for the tensor parameters in the associated mean--field models and, more importantly, for excluding those signs and regions of values that could not provide 
the correct phenomenological trends. 
We proposed three sets for the Skyrme force starting from the existing parametrizations SLy5 \cite{sly5}, 
SIII \cite{siii}, and T$_{41}$ \cite{t41}, and one set for the Gogny force starting from the existing parametrization D1S \cite{d1s}. 
In all cases, also the spin--orbit parameters were modified with respect to the original values. The new parameters are reported in Table III of Ref. \cite{gra2013}. One of these parametrizations was recently employed to analyze the magicity of the nuclei $^{52}$Ca and $^{54}$Ca \cite{gra2014}, and a good agreement with the corresponding experimental findings \cite{nature1,nature2} was found. 

Tensor--induced and pure spin--orbit effects cannot be easily disentangled one from the other, because 
the tensor force has an effect on the spin--orbit splittings in all  
 spin--unsaturated nuclei. Tensor contributions may have for these systems a relevant impact on increasing or reducing such  splittings. 
Recently, one case was investigated  experimentally to isolate effects produced only by the spin--orbit force, without any tensor--induced contributions: $(d,p)$ transfer reactions were performed to extract the neutron $2p$ and $1f$ splittings in the nuclei $^{37}$S and $^{35}$Si \cite{bur2014}. In both nuclei, we expect tensor effects to be practically the same, because the only difference between them  is the filling of the 
proton $2s$ orbital. Consequently, if some difference is measured in the splittings, it may be unambiguously related to a modification of the potential generated by the spin--orbit force. 

Some years ago, a strong charge--density depletion (between 25\% and 30\%) was predicted in the interior of the nucleus $^{34}$Si \cite{gra2009}. This depletion, not predicted in the nucleus $^{36}$S, is simply related to the fact that the proton $2s$ orbital is not occupied in the nucleus     
$^{34}$Si, whereas it is filled in the nucleus $^{36}$S. Since the spin--orbit potential can be written in terms of derivatives of the nuclear densities,  
a strong depletion of the density in the center of the nucleus is expected to have an important impact mostly on the splitting of 
low--$l$ spin--orbit partners, whose 
wave funstions are more concentrated in the internal region. 
This is illustratred in Ref. \cite{bur2014}, that is, a reduction of the neutron $2p$ spin--orbit splitting  from $^{37}$S to $^{35}$Si: 
by using the major fragments to evaluate the spin--orbit splittings, the experimental values reported in Ref. \cite{bur2014} are 2 and 1.1 MeV in the nuclei 
$^{37}$S and $^{35}$Si, respectively, that means a reduction of 45\%. This value was however corrected and reduced after performing shell--model calculations to include correlation effects.  

On the other side, the results of Ref. \cite{bur2014} do not allow us to draw precise conclusions for the 
$f$ splitting: the 1$f_{5/2}$ state 
is represented by three fragments in the nucleus $^{37}$S (with a quite low associated spectroscopic factor equal to 0.36), and is spread in a region of about 1 MeV in the nucleus 
$^{35}$Si (with a low associated spectroscopic factor of 0.32). 

For the $p$ case, this measurement represents a direct and reliable insight into the spin--orbit potential (without any tensor—-induced effects). It is also an indirect confirmation of the theoretical 
prediction of Ref. \cite{gra2009}, about the strong central depletion of the charge density in the nucleus $^{34}$Si. 

In this work, we use (i) the original SLy5 and D1S forces, (ii)  the tensor parametrization of Ref. \cite{colo} for the Skyrme case and one Gogny parametrization of Ref. \cite{angu2012}, and (iii) 
two parametrizations of Ref. \cite{gra2013}, one based on SLy5 and one based on D1S, to analyze pure and tensor--induced spin--orbit effects on the neutron $2p$ and $1f$ splittings in the $N=20$ isotones $^{40}$Ca, $^{36}$S, and $^{34}$Si. 
The case (ii) is an example of 
parametrizations where the spin--orbit parameter is not fitted simultaneously with the tensor parameters,  
whereas in the case (iii) spin--orbit and tensor parameters are adjusted together. 
We perform 
Hartree-Fock (HF) calculations to focus our attention only on pure mean--field effects.  
The article is organized as follows. In Sec. II we present the results and identify pure and tensor--induced contributions. Conclusions are drawn in Sec. III. 

\section{Results. Tensor--induced and pure spin--orbit contributions}

In the adopted mean--field theory, 
two contributions may be identified 
to the single--particle spin--orbit splittings; one depends on the spin--orbit force 
 with a strength related to the spin--orbit parameter of the used effective force; the other is produced by the tensor force with two parameters that describe like--particle and  neutron--proton effects. In the Skyrme case, it is well known that the last contribution is generated by the so--called $J^2$ terms in the Hamiltonian density, where 
$J$ is the spin--orbit density. This contribution  
 is written as the sum of two parts, one related to central--exchange terms (and depending on standard Skyrme parameters), the other describing genuine tensor effects with two parameters. In many Skyrme parametrizations the $J^2$ terms are  neglected in the Hamiltonian density. For the original SLy5 case, the terms coming from the central--exchange contribution are included, whereas the pure tensor terms are neglected.  

Mean--field models cannot describe the fragmentation of single--particle states. Each single--particle state is concentrated at a unique value of the energy. To reasonably compare our energies with the experimental values, we can thus only use the experimental centroids. We remind that in Ref. \cite{bur2014} the shown experimental energies do not correspond to the centroids, but to the main peaks. We cannot thus directly compare our results with them.  
Furthermore, our occupation probabilities are by construction equal to 1 for the occupied states and to 0 for the empty states in HF, and we cannot thus describe any spectroscopic factors. We will concentrate our analysis on the energies of the states and we will compare them with the experimental centroids, in those cases where they are available.  

\begin{figure}
\includegraphics[scale=0.35]{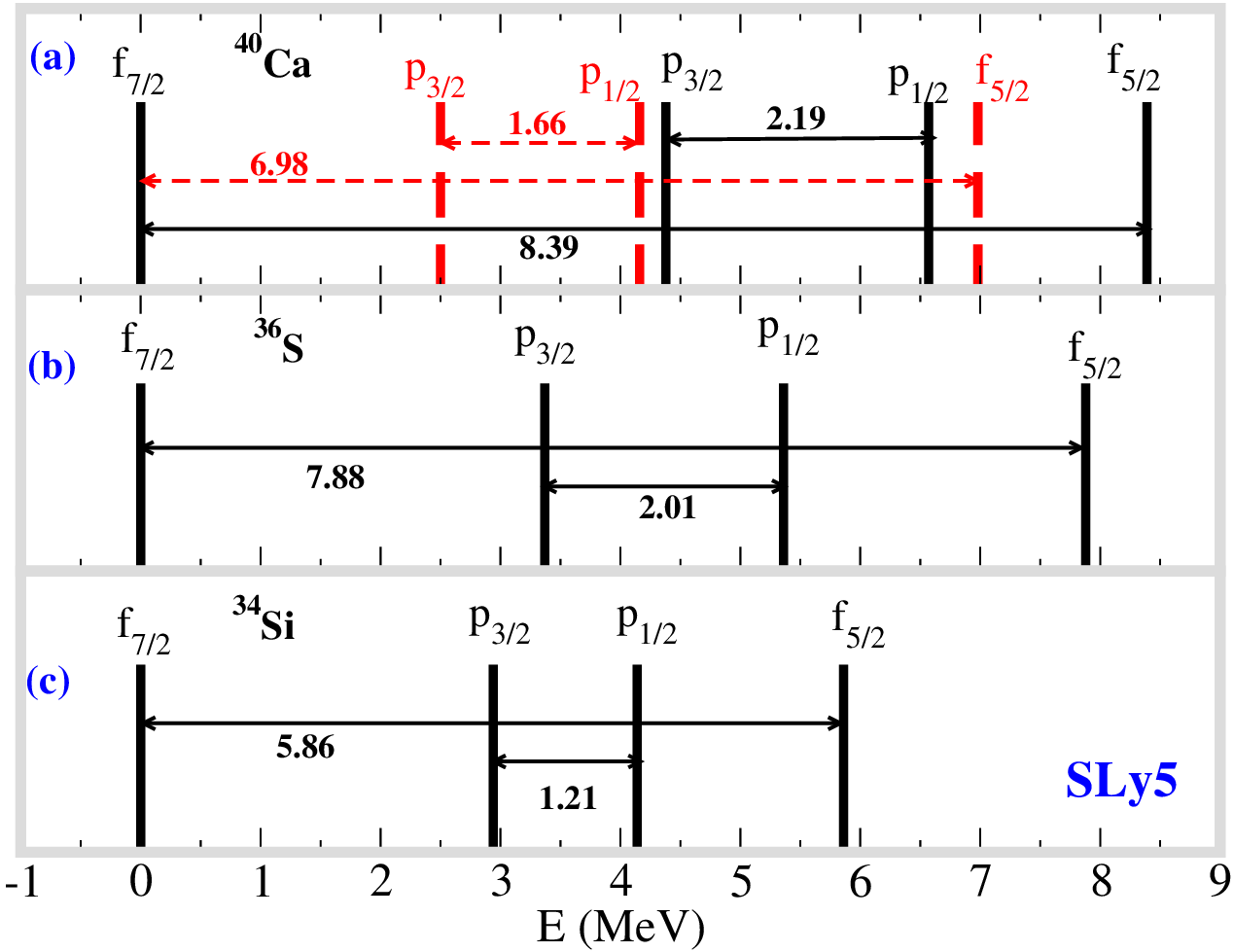}
\caption{(Color online) Energies of the neutron states $2p_{3/2}$, $2p_{1/2}$, and $1f_{5/2}$, with respect to the energy of the neutron state $1f_{7/2}$, located at zero. Panels (a), (b), and (c) refer to the nuclei  $^{40}$Ca, $^{36}$S, and $^{34}$Si, respectively. The calculations are done with the Skyrme parametrization SLy5. The experimental centroids are also plotted in (a) and represented by red dashed lines. The values of the spin--orbit splittings are reported near the corresponding arrows in units of MeV.}
\label{sly5fig}
\end{figure}

\begin{figure}
\includegraphics[scale=0.35]{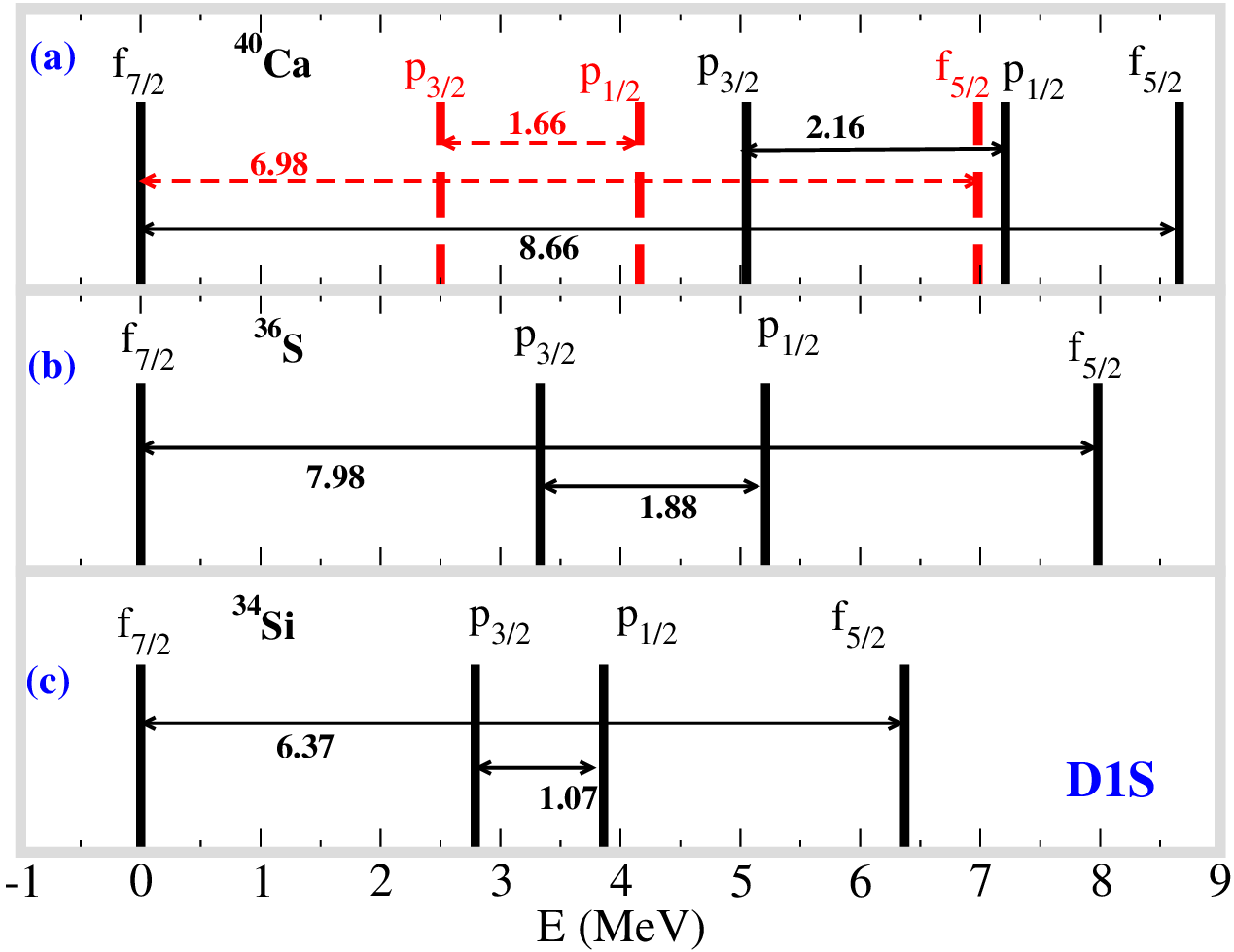}
\caption{(Color online) Same as in Fig. \ref{sly5fig}, but for the Gogny force D1S. }
\label{d1sfig}
\end{figure}

As a first application, we perform calculations by using the original SLy5 and D1S forces. The obtained results for the neutron $2p$ and 
$1f$ spin--orbit splittings are presented in Figs. \ref{sly5fig} and \ref{d1sfig} for the Skyrme and Gogny cases, respectively.  

Panels (a), (b), and (c) refer to $^{40}$Ca, $^{36}$S, and  
$^{34}$Si, respectively. 
The experimental values for the centroids are available only for the nucleus $^{40}$Ca \cite{uoz}. 
They are represented by red dashed lines in Figs. \ref{sly5fig}(a) and \ref{d1sfig}(a).

First, we observe that the  $p$ and $f$ splittings 
calculated for $^{40}$Ca with the Skyrme and Gogny forces 
are too large with respect to the 
experimental values. Going from $^{40}$Ca to $^{36}$S, they are 
weakly reduced. 
Such reductions are not produced by any tensor contributions because such contributions are absent in these calculations. 

A stronger reduction is found for the $p$ splitting going from $^{36}$S to $^{34}$Si, 40\% (43\%) with SLy5 (D1S). 
This important reduction is related to a pure spin--orbit effect, that 
affects much more $p$ than $f$ states  
because $p$ states are more concentrated in the central region of the nucleus where the variation of the
 spin--orbit potential is located, due to the charge--density depletion in the nucleus $^{34}$Si. 
The reduction of the $f$ splitting is smaller, 26\% (20\%) with SLy5 (D1S).  

The state $1f_{5/2}$ in the nucleus $^{36}$S, and the 
states $1f_{5/2}$ and $2p_{1/2}$ in the nucleus $^{34}$Si are not bound in the Skyrme case. In the Gogny case, in addition, the states $2p_{1/2}$ in the nucleus $^{36}$S 
and $2p_{3/2}$ in the nucleus $^{34}$Si are unbound. To estimate their energies within the HF model, we have performed calculations using a box discretization with  an increasing box radius, equal to 20, 40, and 60 fm. 
The estimation of the energies has been done by combining the following criteria: the energies of single--particle resonant states are expected to be weakly 
affected by the modification of the box radius, within their width; these states have wave--function radial profiles similar to those of bound states, 
except in the region far from the nucleus.

\begin{figure}
\includegraphics[scale=0.35]{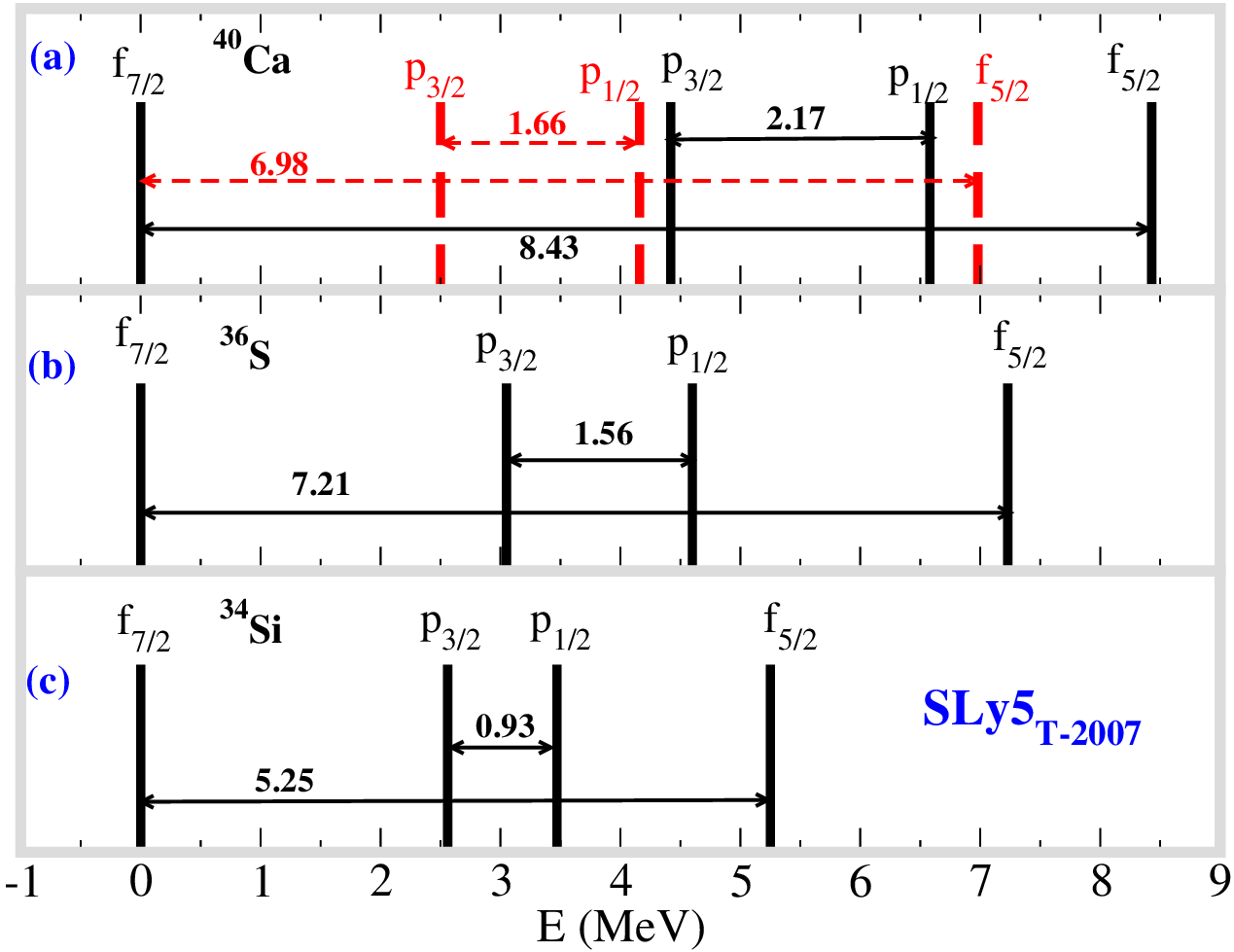}
\caption{(Color online) Same as in Fig. \ref{sly5fig} but with the interaction SLy5$_{T-2007}$.}
\label{sly5tfig}
\end{figure}

\begin{figure}
\includegraphics[scale=0.35]{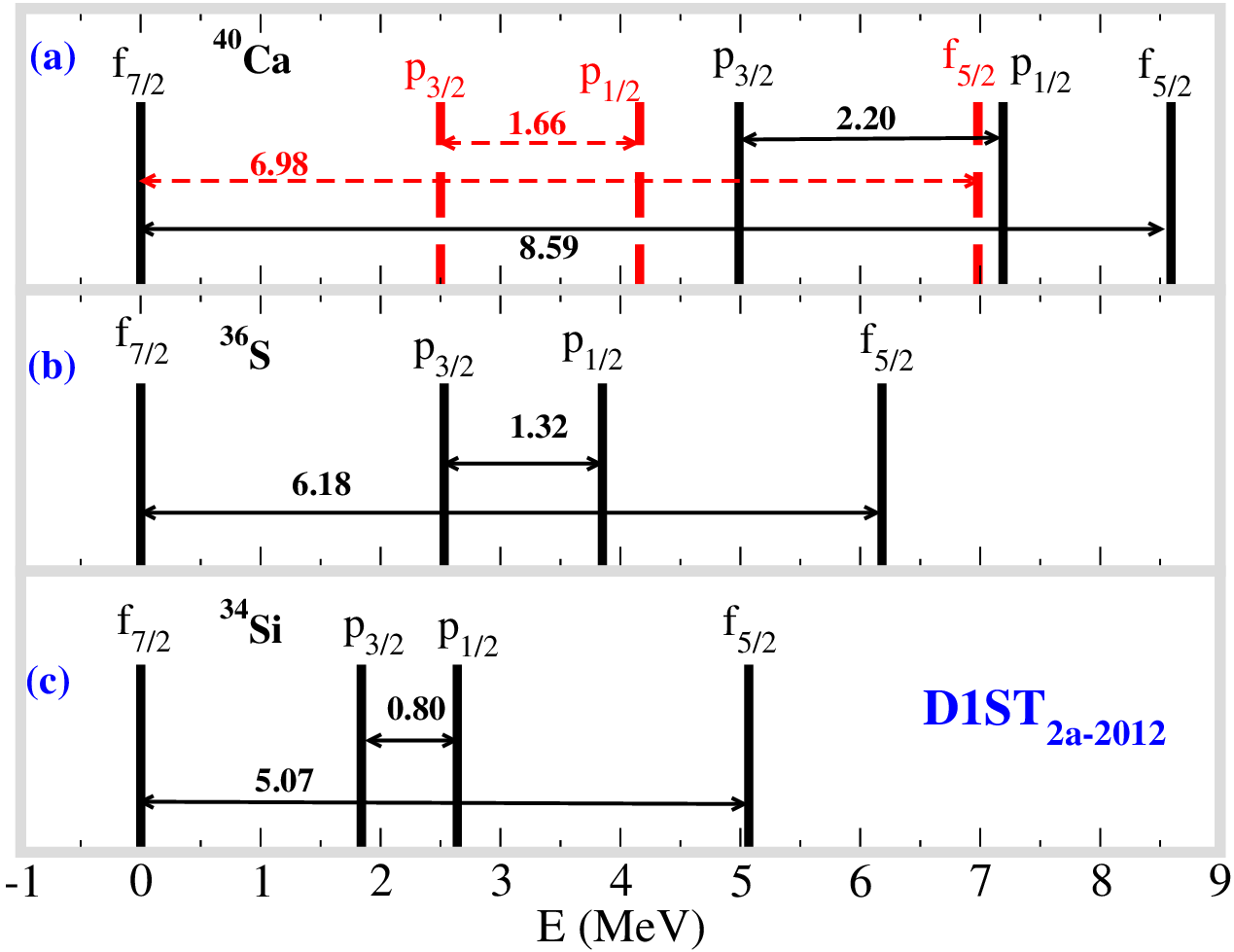}
\caption{(Color online) Same as in Fig. \ref{sly5fig} but with the interaction D1ST$_{2a-2012}$. }
\label{gogny2fig}
\end{figure}

Figures \ref{sly5tfig} and \ref{gogny2fig} show the same results as Figs. \ref{sly5fig} and \ref{d1sfig}, obtained now with the interaction 
of Ref. \cite{colo} and the interaction D1ST2a of Ref. \cite{angu2012}, respectively. We call SLy5$_{T-2007}$ the parametrization of Ref. \cite{colo} and D1ST$_{2a-2012}$ the parametrization D1ST2a. Tensor effects are taken into account, and the spin--orbit parameter is equal to 
that of the original forces. The results found for $^{40}$Ca are almost the same as those obtained in the previous case because this nucleus is spin saturated. 

The splitting reduction found 
from  $^{40}$Ca to $^{36}$S is 
more important than that found in the previous case, because 
neutron--proton tensor effects are now taken into account. The reduction is 14\% (28\%) for the $f$ case and 28\% (40\%) for the $p$ case with SLy5$_{T-2007}$ (D1ST$_{2a-2012}$). The reduction produced by the pure spin--orbit force from 
 $^{36}$S to $^{34}$Si  is  40\% (39\%) for the $p$ case and 
27\% (18\%) for the $f$ case with SLy5$_{T-2007}$ (D1ST$_{2a-2012}$).

Finally, the same calculations have been repeated with two parametrizations of Ref. \cite{gra2013}, where the spin--orbit parameter was 
also modified compared to the original forces. 

\begin{figure}
\includegraphics[scale=0.35]{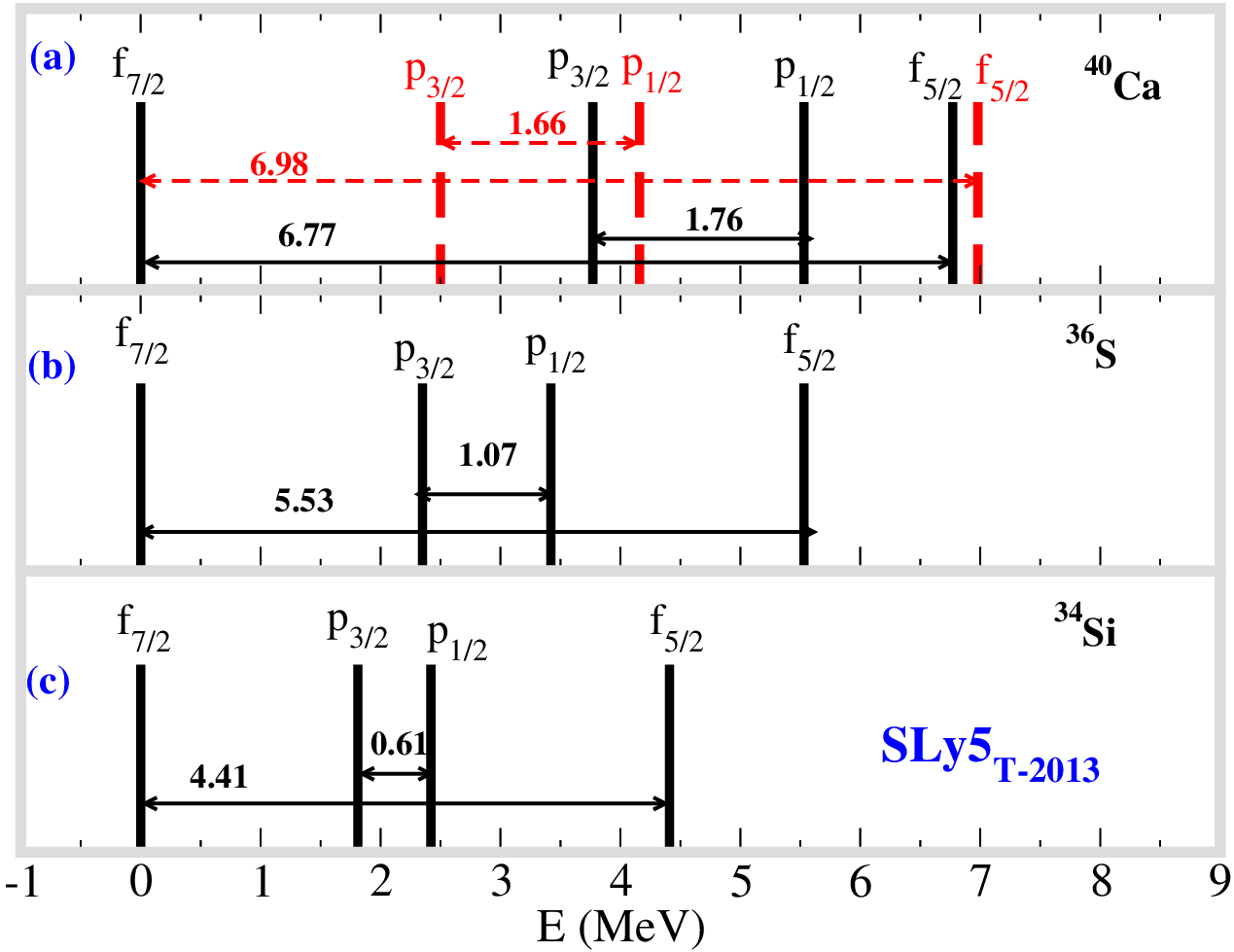}
\caption{(Color online) Same as in Fig. \ref{sly5fig} but with the interaction SLy$_{T-2013}$.}
\label{sly5t2013fig}
\end{figure}

\begin{figure}
\includegraphics[scale=0.35]{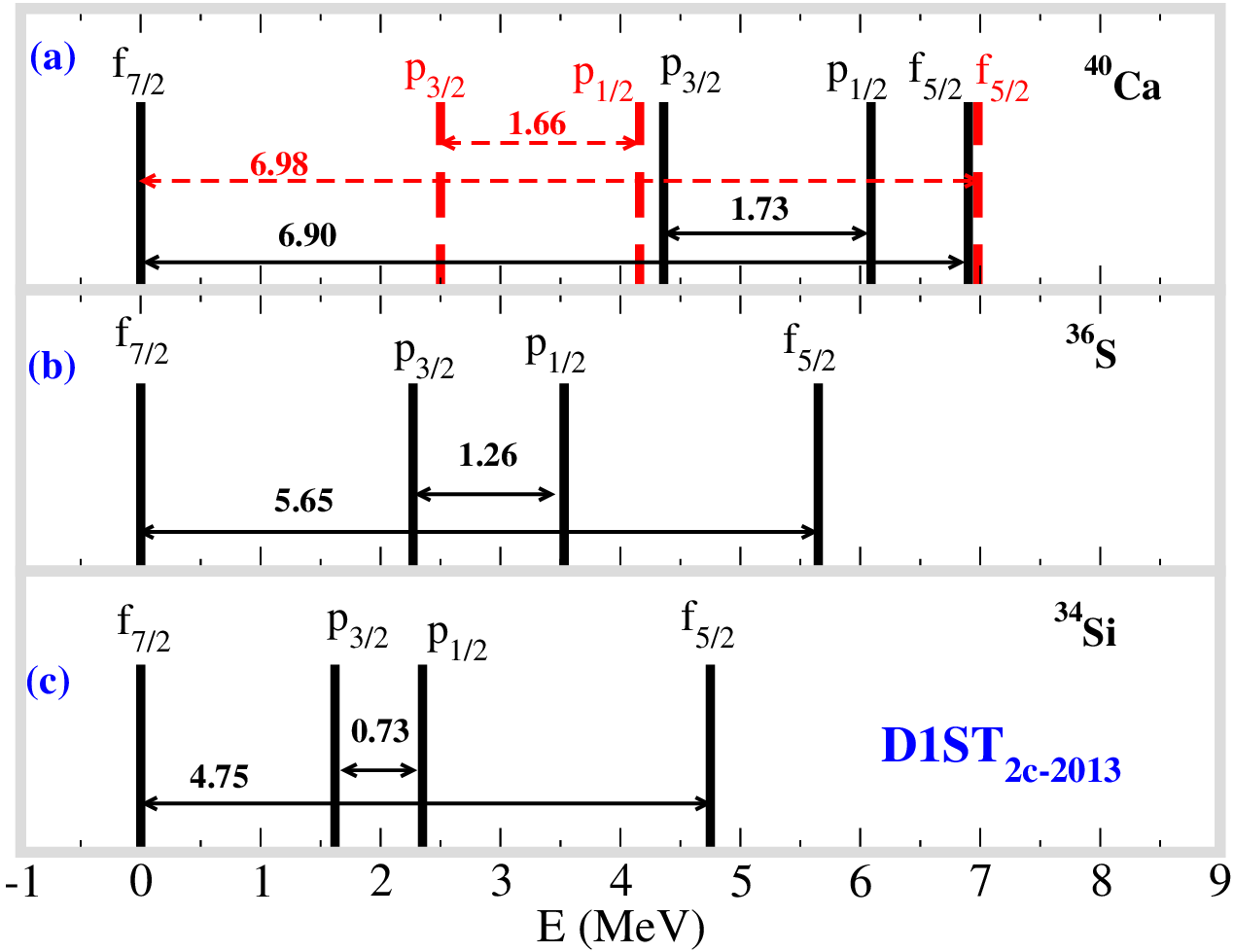}
\caption{(Color online) Same as in Fig. \ref{sly5fig} but with the interaction D1ST$_{2c-2013}$.}
\label{gogny3fig}
\end{figure}

For the Skyrme case, we use the parameter set constructed on top of SLy5. We denote this parametrization with  
SLy5$_{T-2013}$. 
We call D1ST$_{2c-2013}$ the Gogny parametrization, to be coherent with the acronym D1ST2c already used in Ref. 
\cite{viviana}, where this parametrization was employed.  
Figures \ref{sly5t2013fig} and \ref{gogny3fig} show the results for the Skyrme and  Gogny cases, respectively.    
The spin--orbit parameter was fitted for these interactions to provide the correct neutron $1f$ splitting in the nucleus 
$^{40}$Ca. We observe that this modification of the spin--orbit parameter compared to the original force also leads to a $2p$ splitting that is in better agreement with the experimental value, compared to the previous two cases. 

The splittings are reduced going to the nucleus $^{36}$S owing to neutron--proton 
tensor effects: 18\% (18\%) in the $f$ case and 39\% (27\%) in the $p$ case with SLy5$_{T-2013}$ (D1ST$_{2c-2013}$). 
From $^{36}$S to $^{34}$Si, the reductions are 43\% (42\%) and 20\% (16\%) for the $p$ and $f$ splittings, respectively, with 
SLy5$_{T-2013}$ (D1ST$_{2c-2013}$). 

We summarize all the reductions found for the neutron $p$ and $f$ splittings in Table I. 

\begin{widetext}
\begin{center}
\begin {table} 
  \begin{center}
\begin{tabular}{ccccc}
\hline
& From $^{40}$Ca to $^{36}$S & (mostly tensor induced) -  & - From $^{36 }$S to $^{34}$Si & (mostly spin orbit) \\
\hline
\hline
Splitting  & SLy5 & D1S &SLy5 & D1S \\
\hline
\hline
$p$ & 8\%	 &	13\%& 40\% &  43\%  \\             
$f$ &	6\% &	8\% & 26\% &  	20\% \\
\hline
\hline             
& SLy5$_{T-2007}$ & D1ST$_{2a-2012}$ & SLy5$_{T-2007}$ & D1ST$_{2a-2012}$ \\
\hline
\hline
$p$ & 28\%	 &	40\%& 40\% &  39\%  \\             
$f$ &	14\% &	28\% & 27\% &  	18\% \\
\hline
\hline   
& SLy5$_{T-2013}$ & D1ST$_{2c-2013}$ & SLy5$_{T-2013}$ & D1ST$_{2c-2013}$ \\
\hline
\hline
$p$ & 39\%	 &	27\%& 43\% &  42\%  \\             
$f$ &	18\% &	18\% & 20\% &  	16\% \\
\hline   
\end{tabular}\caption {\label{tab-0} Reductions of neutron $p$ and $f$ splittings.}
\end{center}
\end{table}
\end{center}
\end{widetext}

\begin{figure}
\includegraphics[scale=0.35]{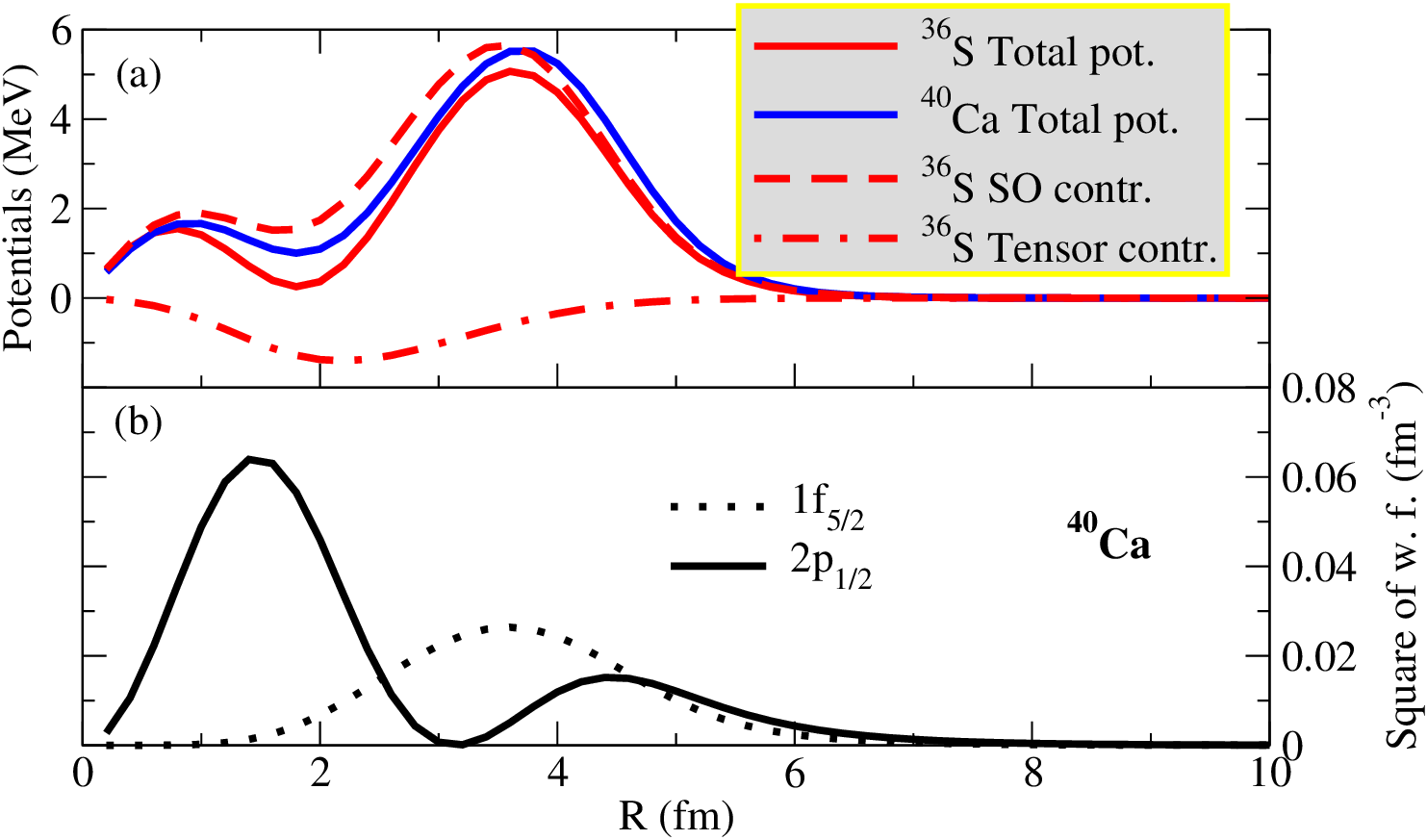}
\caption{(Color online) (a) Total spin--orbit potentials for $^{40}$Ca (solid blue line) and $^{36}$S (solid red line), and 
spin--orbit (red dashed line) and tensor (red dot--dashed line) contributions in $^{36}$S; (b) Squares of the wave functions in the nucleus 
$^{40}$Ca for the $1f_{5/2}$ (dotted line) and the $2p_{1/2}$ neutron states. All the calculations are performed with the interaction SLy5$_{T-2013}$.}
\label{pote}
\end{figure}

First, we observe a general trend. 
In all cases both splitting reductions, the one generated by the neutron--proton tensor 
contribution, from $^{40}$Ca to $^{36}$S, and the one produced by pure spin--orbit effects, from $^{36}$S to $^{34}$Si, are more 
pronounced for $p$ than for $f$ states. 
From  $^{36}$S to $^{34}$Si, 
 this fact was already associated to the modification of the spin--orbit 
potential in the center of the nucleus $^{34}$Si, with a more important overlap with $p$ than with $f$  wave functions.
The reduction from $^{40}$Ca to $^{36}$S, on the other side, is related to neutron--proton tensor effects (the proton $1d_{3/2}$ orbital is filled in 
$^{40}$Ca and empty in $^{36}$S). 
As an illustration, we analyze the nuclei $^{40}$Ca and $^{36}$S with the parametrization  SLy5$_{T-2013}$. It is known that 
the spin--orbit potential in the Skyrme--mean--field approach is expressed as the sum of two terms; one is produced by 
the spin--orbit force, depends on the derivatives of the densities, and is tuned by the spin--orbit parameter; the second one 
is induced by the tensor force, depends on the spin--orbit densities, and is tuned by the tensor 
parameters. We plot in the upper panel of Fig. 
\ref{pote} some spin--orbit potentials for the two nuclei. For $^{40}$Ca, only the total spin--orbit potential is shown because it coincides with the pure spin--orbit contribution, the tensor contribution being negligible in this saturated nucleus. 
For $^{36}$S, we display the total potential, together with the spin--orbit and tensor contributions. For this nucleus, the 
tensor--induced contribution is produced by a spin--orbit density constructed predominantly with the proton $d$ wave function. This 
contribution has the opposite sign compared to that coming from the spin--orbit force, and produces a total 
potential which is reduced with respect to the pure spin--orbit part. The volume localization of such reduction is related to the 
localization of the $d$ wave function. The net result is that the total potential for the nucleus $^{36}$S is lower than that 
for the nucleus $^{40}$Ca.

In the lower panel of the figure, we show the squares of the wave functions for the neutron states $1f_{5/2}$ and 
$2p_{1/2}$, as an illustration, in the nucleus $^{40}$Ca.
It turns out that the two total spin--orbit potentials (for $^{40}$Ca and $^{36}$S) 
have a stronger overlap 
with the $f$ wave function than with the $p$ wave function. Owing to the reduction of the potential 
from $^{40}$Ca to $^{36}$S, both overlaps are decreased generating a reduction of both splittings from $^{40}$Ca to $^{36}$S. 
However, the  
 overlap reduction is more important for the $p$ (22\%) than for the $f$ wave function (13\%). This explains 
why 
the splitting is predicted more strongly reduced for $p$ than for $f$ states. 
It is clear that this prediction depends in our model on the sign of the tensor parameter governing the neutron--proton contribution. With a different sign of the corresponding tensor parameter, the total spin--orbit potential would be increased instead of reduced passing from $^{40}$Ca to $^{36}$S, and both  splittings would be larger in $^{36}$S than in $^{40}$Ca.  

Let us concentrate now on the splitting reductions related to tensor--induced effects, that is,  from $^{40}$Ca to $^{36}$S. In this case, the reductions are very small with the interactions D1S and SLy5 because the tensor force is absent there. 
We observe that both $p$ and $f$ splittings are increased in the Skyrme case, passing from SLy$_{T-2007}$ to SLy$_{T-2013}$, 
and reduced in the Gogny case, passing from D1ST$_{2a-2012}$ to D1ST$_{2c-2013}$. This occurs because the parameter 
governing the neutron--proton tensor contribution is larger in  SLy$_{T-2013}$ compared to SLy$_{T-2007}$, and 
smaller in  D1ST$_{2c-2013}$ compared to D1ST$_{2a-2012}$. 

For the reduction of the splitting related to the pure spin--orbit force, from $^{36}$S to $^{34}$Si, we have found in all cases
 a reduction of about 40\% for $p$ states and of about 20\% for $f$ states. These reductions do not seem strongly 
affected by the specific values of the parameters.  Even if such reductions are predicted to be almost the same with all the employed interactions, we observe that only for the last case, SLy$_{T-2013}$ and D1ST$_{2c-2013}$, the starting point is correct, that is, the splittings are correctly reproduced in the nucleus $^{40}$Ca with respect to the experimental values. 

The 40\% reduction of the $p$ splitting is  
comparable to the experimental finding (before correction for correlations) even if such comparison is not 
completely meaningful because the experimental energies refer to the major fragments. 
It would be interesting to have all the experimental values for the centroids in the nuclei  $^{37}$S and $^{35}$Si, to compare with our results. 
Furthermore, having the centroids for the $f$ states would 
permit to validate our prediction on the stronger reduction of the 
 $p$ (compared to the $f$) splitting in all cases, from $^{40}$Ca to 
$^{36}$S (mostly tensor effects), and from $^{36}$S to $^{34}$Si (mostly pure spin--orbit effects).

\section{Conclusions}
We have computed neutron $2p$ and $1f$ spin--orbit splittings in the isotones $^{40}$Ca, $^{36}$S, and $^{34}$Si, by 
using mean--field approaches based on Skyrme and Gogny forces. We have used three types of interactions, (i) 
without tensor contributions, (ii) with tensor contributions and with tensor parameters adjusted on top of existing parametrizations, (iii) with the tensor force and with tensor parameters adjusted together with the spin--orbit parameter. 
Our conclusions can be directly drawn from Table I. For all the interactions, we predict a reduction 
of both splittings passing from $^{40}$Ca to 
$^{36}$S and from $^{36}$S to $^{34}$Si. From 
$^{40}$Ca to 
$^{36}$S, the reduction of the splittings is related to neutron--proton tensor--induced effects and can be easily 
understood by analyzing the strengths of the parameters tuning the neutron--proton tensor contribution in each case. This  explains  why 
this reduction is very small for the interactions of type (i), where the tensor force is not included. 
This reduction is stronger for the $p$ splitting than for the $f$ splitting and the reasons for it, related to the overlap modifications between the spin--orbit potentials and the wave functions, were analyzed in Sec. II. From  
$^{36}$S to $^{34}$Si, the reduction is predominantly coming from the spin--orbit force. This reduction is of about 40\% for the $p$ case and of about 20\% for the $f$ case. Again, the $p$ reduction is stronger and this can be easily understood as due to 
the central charge--density depletion in the nucleus $^{34}$Si. This depletion is concentrated in the center of the nucleus where $p$ wave functions are more localized than $f$ wave functions. In our model, the predicted $p$ reduction is of about 40\%. This reduction cannot be directly 
compared with the value reported in Ref. \cite{bur2014} because such value is obtained by using the main fragments and not the centroids. 
We stress once again that the comparison between centroids and major fragments may be indeed quite misleading in those cases where such values are very different one from the other, owing to a strong fragmentation of the strength. This was shown for instance in the experimental results presented in Ref. \cite{eckle}. 
The measurement of the centroids in $^{36}$S and $^{34}$Si would provide values that we could more properly compare with our results. 
In particular, one could verify our prediction on the stronger $2p$ reductions (both from  $^{40}$Ca to 
$^{36}$S and from $^{36}$S to $^{34}$Si), compared to the $1f$ reductions.

%
%
%

\end{document}